\documentclass{uncecomp2017}
\usepackage{amsmath}
\usepackage{subcaption}

\title{Mutliscale Surrogate Modeling and Uncertainty Quantification for Periodic Composite Structures}

\author{Charilaos Mylonas$^1$,Bemetz Valentin$^1$, Eleni Chatzi$^1$}

\heading{Charilaos Mylonas, Valentin Bemetz and Eleni Chatzi}

\address{$^1$
Department of Civil, Environmental \\ and Geomatic Engineering, \\ ETH Z\"urich \\ Stefano-Franscini-Platz 5, 8093 Z\"urich\\
  e-mail: mylonasc@ibk.baug.ethz.ch
  }


\abstract{ Computational modeling of the structural behavior of continuous fiber composite materials 
often takes into account the periodicity of the underlying micro-structure. A well established method 
dealing with the structural behavior of periodic micro-structures is the so-called \textit{Asymptotic Expansion Homogenization} (AEH).
By considering a periodic perturbation of the material displacement, scale bridging functions, also referred to as \textit{elastic correctors},
can be derived in order to connect the strains at the level of the macro-structure with micro-structural strains. 
For complicated inhomogeneous micro-structures, the derivation of such functions is usually performed by the numerical solution of a PDE problem - typically with the Finite Element Method.
Moreover, when dealing with uncertain micro-structural geometry and material parameters, there is considerable uncertainty introduced in the actual 
stresses experienced by the materials. Due to the high computational cost of computing the elastic correctors, the choice of a pure Monte-Carlo approach 
for dealing with the inevitable material and geometric uncertainties is clearly computationally intractable. This problem is even more pronounced 
when the effect of damage in the micro-scale is considered, where re-evaluation of the micro-structural representative volume element is necessary for every occurring damage.
The novelty in this paper is that a non-intrusive surrogate modeling approach is employed with the purpose of directly bridging the macro-scale behavior of 
the structure with the material behavior in the micro-scale, therefore reducing the number of costly evaluations of corrector functions, allowing for 
future developments on the incorporation of fatigue or static damage in the analysis of composite structural components. }

\begin{document}

\section{Introduction}
Continuous fiber reinforced 
polymer composites are light, 
stiff materials of significantly improved static strength and fatigue resistance.
 For the engineering analysis of such materials, a direct 
 discretization of the fine 
 spatial variation of the composite material
would render the problem computationally intractable.
Therefore, composite engineering analysis seeks to deliver
a consistent calculation of the effective macroscopic properties 
by considering the material and geometrical properties of 
the micro-structure (\textit{homogenization}) and 
adequately approximating the stresses in the micro-structure (\textit{localization\footnote{Not to be confused with 
\textit{localization} in the context of damage detection. Some authors use the term \textit{de-homogenization} to avoid
confusion.}}).
A mathematically rigorous approach to the problem of homogenization 
and localization, that further applies to the problem of elasticity 
in the context of composite materials, was 
proposed in \cite{homogenization78}.

The aforementioned technique is often termed 
\textit{Asymptotic Expansion Homogenization} (AEH). 
The application of the method relies on 
the assumption that displacement appears into well 
separated spatial scales. The method yields effective elastic 
properties on the macro-scale without any assumptions on 
the distribution of strains or stresses in the micro-scale, 
but only with the assumption of periodicity in the 
displacements among different representative volume elements.
For a more detailed, engineering oriented derivation of the 
AEH method for elasticity the reader is referred to \cite{chung2001} and \cite{PinhodaCruz}.
Extending the AEH method to damaged composites has also attracted
research interest \cite{oskay2007eigendeformation,michel2003nonuniform}.
In the context of another homogenization framework, 
it has been shown that efficient hysteretic multi-scale 
damage models can be derived 
\cite{triantafyllou2014hysteretic}.

The complete determination of material 
properties and microstructure geometry is,
in general, not possible. Therefore, the 
prediction of the material response in the micro-scale 
should account for uncertainty.
A direct Monte-Carlo approach 
for the purpose of representing 
the effect of all the uncertain 
parameters would quickly become intractable.

In the present work we investigate the potential of 
non-intrusive probabilistic uncertainty propagation techniques, 
namely the Polynomial Chaos Expansion (PCE)\cite{xiu2002wiener},
for the purpose of constructing surrogate models.
Efficient surrogate modelling techniques are expected to yield further 
reductions in the computational cost of multi-scale finite element analysis.
An intrusive PCE for the same problem was proposed in \cite{homogPCE}.

Finally, a dimensionality reduction technique, namely Principal Component Analysis (PCA)
was found to be highly efficient on decomposing the stiffness tensor without strong 
assumptions on the geometry induced symmetries of the homogenized stiffness tensor.

The uncertainty quantification toolbox UQLab was used for deriving 
the PCE of the homogenized stiffness tensor \cite{uqlabref}.

\section{Computational Methodology}
In the following the basic components of 
the Asymptotic Expansion Homogenization 
for analyzing elastic periodic structures 
and the Polynomial Chaos Expansion surrogate 
modelling technique, used in the present study are going
to be briefly presented.

\subsection{Asymptotic Expansion Homogenization}
This section serves for establishing notation
notation and introducing an intuitive understanding 
of the quantities related to the problem of homogenization and 
localization for periodic media.
We denote  $\mathbf{x}= \{x_1,x_2,x_3\} $ as the coordinate 
system of a composite structure, and further introduce a coordinate system local to 
every representative volume element (microstructure) $\mathbf{y} = \{ y_1,y_2,y_3 \}$.
Quantities marked with $\cdot^\epsilon$ 
denote the high resolution quantities in the macro-scale.
All indices attain values in $\{1,2,3\}$. Einstein summation is implied for repeating indices.
We seek to solve the elasticity boundary value problem,
\begin{align}
	\label{eq:elasticity}
	\frac{\partial{\sigma^\epsilon_{ij}}}{\partial x^\epsilon_{j}}+f_i & = 0  \quad    & in & \quad \Omega \\
	u^\epsilon_i                                                       & = 0  \quad    & on & \quad \partial_1 \Omega \\ 
	\sigma^{\epsilon}_{ij} n_{j}  & = F_i                                               & on & \quad \partial_2 \Omega \\ 
	\epsilon_{ij} ( \mathbf{u^\epsilon} ) & 
	        = \frac{1}{2} \big( \frac{\partial u^\epsilon_i}{\partial x^\epsilon_j}  + \frac{\partial u^\epsilon_j}{\partial x^\epsilon_i} \big) \\
\end{align}
where $\partial_1 \Omega $ and $\partial_2 \Omega$ denote
different boundaries, $u^\epsilon_i = \mathbf{u}^\epsilon(\mathbf{x})$ is 
the displacement of the macro-structure, $F_i$ a traction force, and $f_i$ 
the body force.
The constitutive relation simply reads 
\begin{equation}
	\sigma^\epsilon_{ij} = D^\epsilon_{ijkl} \epsilon_{kl}.
\end{equation}
Due to the geometry of the continuous fiber reinforced 
composites, $D^\epsilon_{ijkl}$ is varying periodically 
in the material, in a scale much finer than the 
scale of the structure.
It is convenient to define the so-called 
\textit{scale}  parameter $\epsilon << 1 $,
which represents the ratio between 
the microscopic and macroscopic scale. 
Considering the coordinates of the micro-scale and 
the macroscale, one may write  $y_i = x_i / \epsilon$.
By the chain rule we have
\begin{equation}
	\frac{\partial \cdot}{\partial x^\epsilon_i} = \frac{\partial \cdot}{\partial x_i} +  \frac{1}{\epsilon} \frac{\partial \cdot}{\partial y_i}
\end{equation}
The displacements are represented with the following 
expansion in, $\epsilon$, as
\begin{equation}
	\label{eq:asexp}
	\mathbf{u}^\epsilon_i(\mathbf{x}) = \mathbf{u}^{(0)}(\mathbf{x}) +  \epsilon \mathbf{u}^{(1)} (\mathbf{x})  +\epsilon^2\mathbf{u}^{(2)} (\mathbf{x}) + \cdots 
\end{equation}
It has been rigorously established \cite{homogenization78}, 
that by plugging \autoref{eq:asexp} into the problem of elasticity, 
and by passing to the limit $\epsilon \rightarrow 0 $,
the elasticity problem boils down to a hierarchical set 
of partial differential equations. 
It is assumed that the displacements in the representative volume elements
are connected to the gradients of the displacement 
in the macro-scale $\frac{\partial u^{(0)}_k}{\partial x_l} (\mathbf{x}) $ by a certain vector valued function $\chi^{kl}_i(\mathbf{y})$. This approximation reads
\begin{equation}
	\label{eq:globloc}
	u_i^{(1)}(\mathbf{x},\mathbf{y}) = -\chi^{kl}_i(\mathbf{y}) \frac{\partial u^{(0)}_k}{\partial x_l} (\mathbf{x}) + \bar u^{(1)}_i(\mathbf{x}),
\end{equation}
where  $\bar u^{(1)}_i(\mathbf{x}) $ denotes the average displacement 
of the representative unit cell in the macro-scale coordinate system. 
Function $\chi^{mn}_i$ is often termed the 
\textit{elastic corrector}.
Note that every pair of components $mn$ correspond to a different 
spatial gradient. The accuracy of this approximation relies on the existence of the gradients 
$\frac{\partial u^{(0)}_k}{\partial x_l} (\mathbf{x}) $
and assumes a slow variation in the macroscopic scale.

For continuous fiber composites, without stress concentrations
this is a reasonable assumption.
A stress concentration may be due to localized damage,
i.e., due to a macroscopic crack or very close to the boundaries 
of the composite structure
\footnote{On the other hand, the effect of 
diffuse slowly spatially varying damage may be well 
approximated without the presented framework to 
break down.}.

For a first order (first order perturbation) approximation of the perturbed displacement field, 
assuming $\chi^{mn}_i$ smooth in $\Omega$ and smooth and periodic 
with zero mean in the RVE or $  \in \mathcal V^{per}$ on $\Omega_Y$, 
the variational problem
\begin{equation}
	\label{eq:corrvar}
	\int_{\Omega_Y} D_{ijkl} \frac{\partial \chi^{mn}_k}{\partial y_l} \frac{\partial \nu_i}{\partial y_j} d \mathbf{y} = \int_{\Omega_Y} D_{ijmn}\frac{\partial \nu_j }{\partial y_i} d \mathbf{y}
\end{equation}
holds. We seek solutions for $\chi^{mn}_i$ so that 
\autoref{eq:corrvar} holds for all $ \nu_i \in \mathcal V^{per}$. 

Due to the symmetries of the stiffness tensor, we have $D_{ijmn} = D_{ijnm} = D_{jimn}$.
Therefore, we only need to consider $ mn = \{11,22,33,23,13,12\} $ for the full computation of the elastic corrector.
In practice \autoref{eq:corrvar} results in 6 variational problems for the computation of the corrector, 
one for every different value of $mn$.

The variational problem allows for a finite element approximation of the
corrector function. By considering the RVE averaged strains and
stresses, an approximation of the stiffness tensor
$D^{\epsilon}_{ijkl} \approx D^{h}_{ijkl}$ 
in the macro-scale is possible. Namely, 
\begin{equation}
	D^h_{ijkl} = \frac{1}{|Y|} \int_{\Omega_Y} D_{ijkl}(\mathbf{y}) \big[ \delta_{kl}\delta_{ln} - \frac{ \partial \chi^{mn}_k}{\partial y_l}  \big] d\mathbf{y}
\end{equation}

In practice, even for the case of homogeneous materials described by Lam\'e parameters in the micro-scale, the
homogenized stiffness tensor turns out anisotropic. Some symmetries may be induced by the geometry,
such as orthotropy and transverse isotropy, but the framework presented in the present work 
is concerned with the case of the fully anisotropic material. 

It is apparent that since the corrector connects the displacements of 
the macro-structure to the displacements of the micro-structure, strains and stresses 
can be straight-forwardly computed for the micro-structure. Namely the micro-stresses are computed with 
\begin{align}
	\label{eq:stresslocl}
	\sigma^{(1)}_{ij}(\mathbf{x}) = D_{ijkl}\big( \delta_{mk}\delta_{nl} - \frac{\partial \chi^{mn}_k}{\partial y_l} \big) \frac{\partial u^{(0)}}{\partial x_n}.
\end{align}

Therefore by storing the solution of the corrector we may directly compute stresses 
in the micro-scale without making any strong assumptions on the distribution of 
stresses or strains on the boundaries of the RVE. The only assumption 
required for this framework is the periodicity of displacements in the boundaries of the RVE.

For the actual solution of the finite element discretization of \autoref{eq:corrvar}, periodic boundary conditions have to be enforced.
In addition, one arbitrary point must be constrained to zero in all components of $\chi^{mn}_i$ since the weak form has a unique solution 
up to an additive constant. Due to the periodicity of the corrector, and the fact that homogenization and localization problems are concerned only with 
derivatives of the corrector, the boundary conditions are essentially equivalent to the zero-mean requirement for the corrector function.

\subsection{Polynomial Chaos Expansions}
Polynomial chaos expansions (PCE) were first introduced in \cite{wiener} for Gaussian input variables and generalized in \cite{xiu2002wiener} for classical probability distribution functions.
Consider a set of random inputs $\mathbf{X} = \{x_1,x_2,\cdots,x_n\} $ to a deterministic model $Y = \mathcal{M}(\mathbf{X})$
The method relies in the construction of a tensor product basis of univariate polynomials $\Phi^{(n)}(x_n)$, 
orthogonal with respect to inner products weighted by probability distribution functions $f_X(x_n)$.
The orthogonality relation reads,
\begin{equation}
	\langle \Phi^{(m)}_i,\Phi^{(m)}_j \rangle_{f_X} = \delta_{ij}
\end{equation}
where 
\begin{equation}
\langle f,g \rangle_{f_X} = \int f(x)g(x)f_{X}(x) dx
\end{equation}
and $\delta_{ij}$ is the Kroneker delta.
The tensor product basis set reads,
\begin{equation}
	\mathbf{\Psi}(\mathbf{X}) = \otimes_{m=1}^{n} \mathbf{\Phi}^{(m)}
	\label{eq:tensprodbasis}
\end{equation}
where $\mathbf{\Phi^{(m)} } = \{ \Phi^{(m)}_1, \Phi^{(m)}_2, \cdots \} $ with superscript denoting the input dimension and subscript denoting the order of the orthogonal polynomial.

A PCE model, is a linear combination of the elements of \autoref{eq:tensprodbasis},
\begin{equation}
	\mathcal{M}(\mathbf{X}) = \sum_{\mathbf{a} \in \mathcal{A}} c_{\mathbf{a}} \mathbf{\Psi}_{\mathbf{a}}(\mathbf{X})
\end{equation}
indexed by $ \mathbf{a} = \{a_1,a_2,\cdots,a_m \} $, which is a multi-index that denotes the degree of the univariate polynomials of each of the input variables, and $\mathcal{A}$ the set of multi-indices.
For example, 
\begin{equation}
	\mathbf{\Psi}_{\mathbf{a}} = \Phi^{(1)}_{a_{1}}(x_1) \Psi^{(2)}_{a_2}(x_2) \cdots \Psi^{(n)}_{a_n}(x_n)
\end{equation}
where $a_i$ denotes the degree of orthogonal polynomials along dimension $i$.
In the presented case the number of random input dimensions is $n=6$.
In practice, the set of multi-indices, is truncated for numerical implementation. Also the PCE is 
considered up to a certain degree of univariate polynomials in each dimension to render the problem 
numerically tractable. According to the Cameron-Martin theorem 
\cite{cameron1947orthogonal,xiu2002wiener}, such an expansion 
converges in the $L_2$ sense, when $\mathcal{M}(\mathbf{X})$ has finite variance.

There are several approaches for the purpose of determining the coefficients 
$c_{\mathbf{a}}$. The most versatile method, that also deals automatically 
with adaptively selecting basis elements, is the 
\text{Least Angle Regression} (LAR)\cite{efron2004least} approach.
LAR is the method of choice for the present work. See \cite{blatmanSudret} and \cite{spiridonakos2016polynomial} for a discussion of the benefits of LAR.

\section{Example application on continuous fibre reinforced composites}
A typical composite structure, composed of transversely isotropic
glass fibers embedded in a polymer matrix with stacking 
sequence $[ 0,-\phi,+\phi] $, was analyzed as a proof of concept.
The material properties adopted herein, are given in \autoref{tab:matprops}.
In this study, only geometric variation of the micro-structure
was considered. For the purpose of demonstrating the 
effectiveness of the PCE surrogate model of the homogenization process,
relatively large variations on the geometrical parameters of the micro-structure were chosen.
The ranges of the parameters chosen for the present work are given in \autoref{tab:geomprops}.

\begin{table}[!h]
	\makebox[0pt][c]{\parbox{1.2\textwidth}}{%
		\centering
\begin{minipage}{0.45\hsize}
	\centering
\begin{tabular}{c|c|c }
	& Fiber & Matrix \\
	\hline 
	$ E_1 $ [GPa] & 31 & 2.79 \\
	$E_2$ [GPa] & 7.59 & 2.76 \\
	$\nu$ & 0.3 & 0.3 \\
	$G_{12}$ & 3.52 & 1.1 \\
	$G_{23}$ & 2.69 & 1.1 \\
\end{tabular}
\caption{Material properties of the micro-structure.}
\label{tab:matprops}
\end{minipage}
\hfill
\begin{minipage}{0.45\hsize}
	\centering
\begin{tabular}{c|c|c}
Parameter               & min & max  \\
\hline
$V_{f_2}$               & 0.600 & 0.74 \\
	${V_{f_1}} /\ { V_{f_2}} $ & 0.600 & 1.00    \\
$a_2$                   & 0.450 & 0.55 \\
	$ {a_1 } /\ { b_1} $        & 0.167 & 0.250 \\
	${a_2}/\ { b_2} $        & 0.167 & 0.250 \\
$\phi $                 & 15    & 75 \\
\end{tabular}
\caption{Assumed micro-structure geometry parameter variations.}
\label{tab:geomprops}
\end{minipage}%
}
\end{table}

\begin{figure}
	\begin{subfigure}{0.5\textwidth}
	\includegraphics[width=0.9\linewidth]{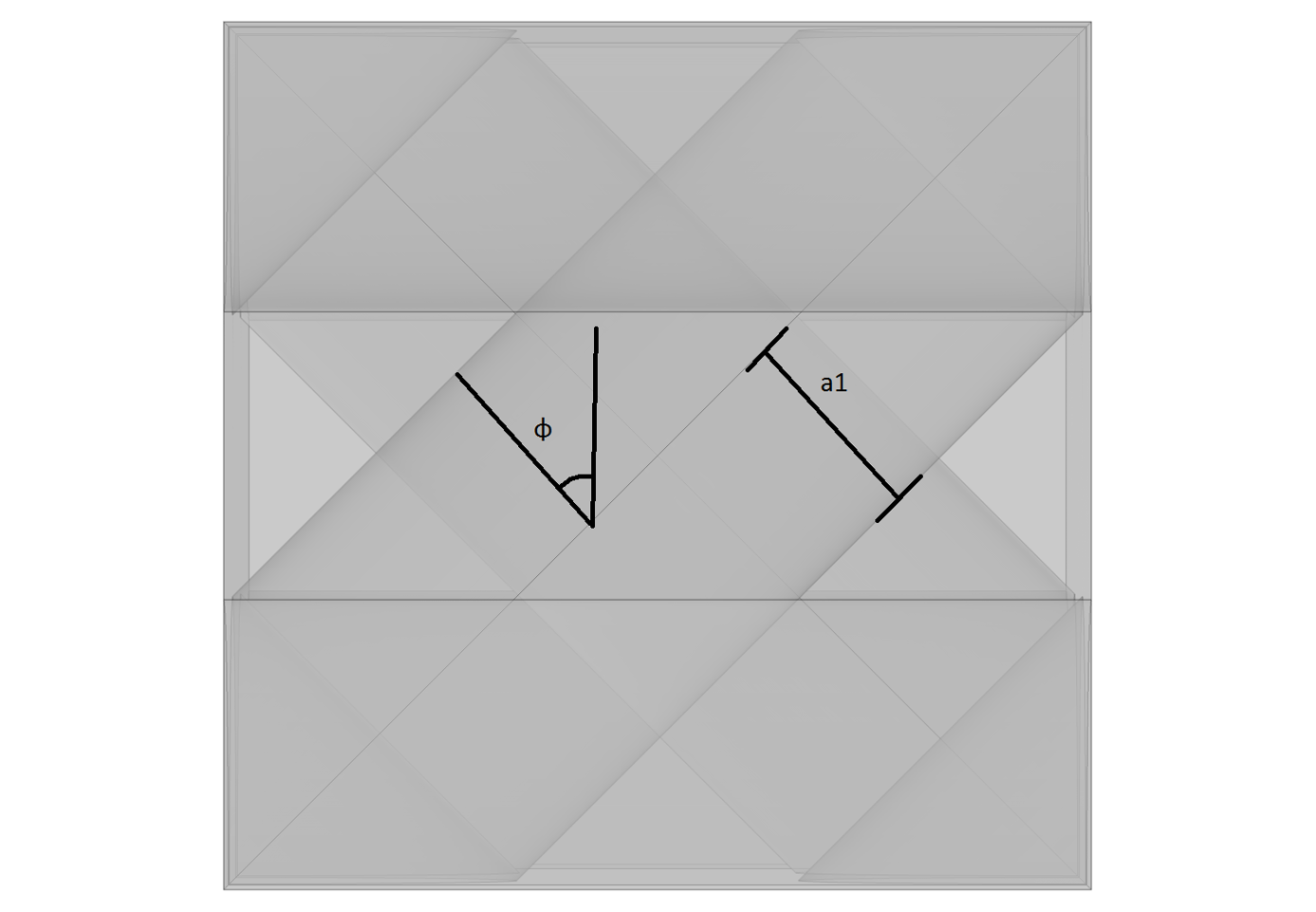}
\end{subfigure}
	\begin{subfigure}{0.5\textwidth}
	\includegraphics[width=0.9\linewidth]{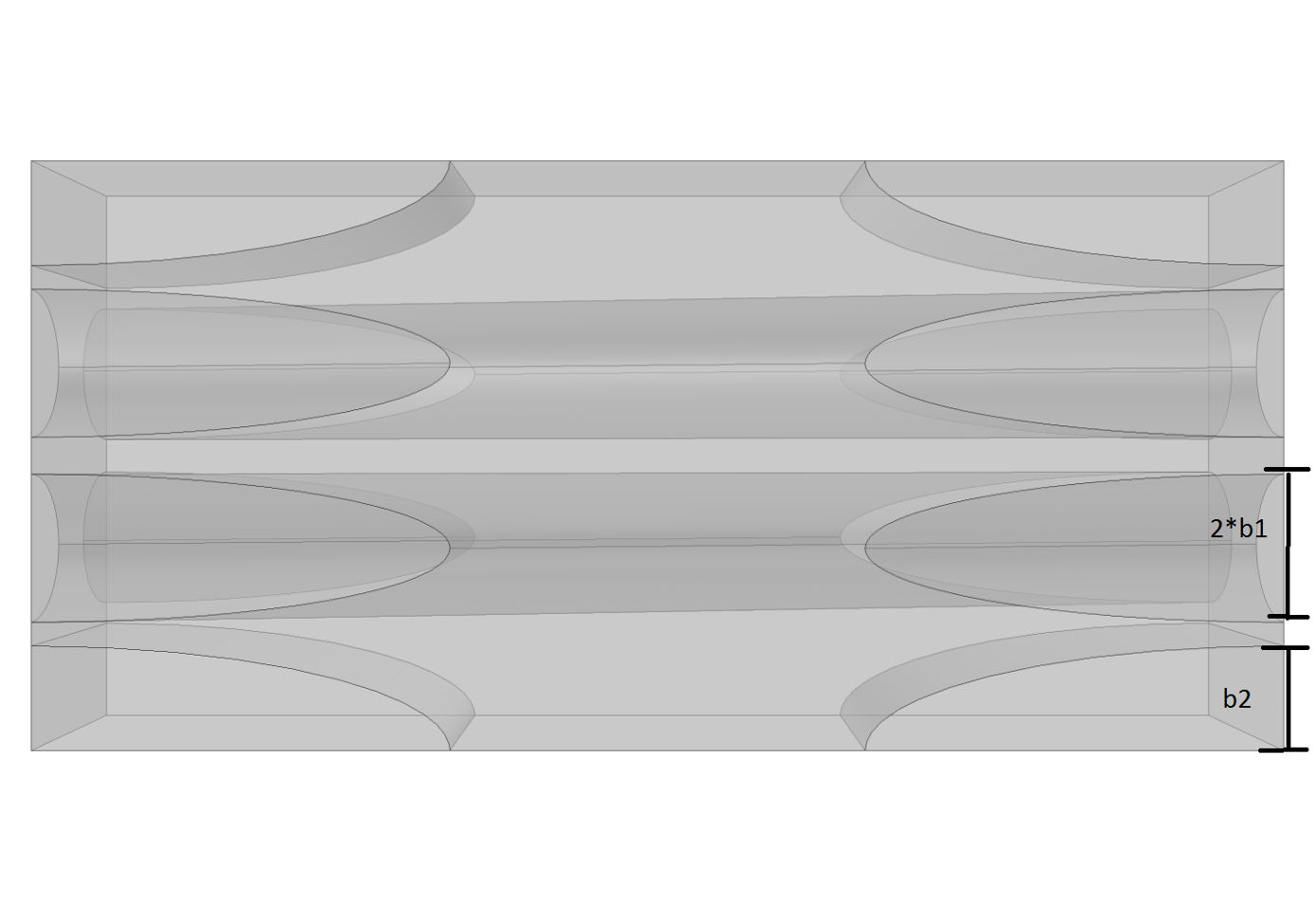}
\end{subfigure}
	\caption{Random geometric parameters of the micro-structure. The volume fractions affect the intra-fiber spacing.}
\label{fig:parametrization}
\end{figure}

\begin{figure}[!h]
\begin{subfigure}[b]{0.5\linewidth}
	\centering
	\includegraphics[width=0.95\linewidth]{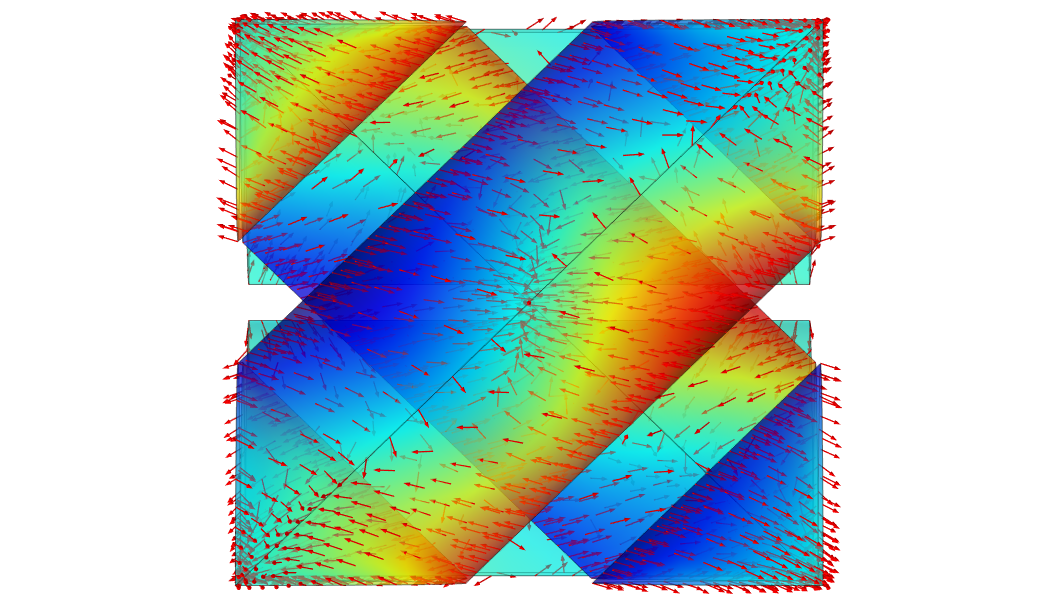}
	\caption{$\mathbf{\chi}^{11}$}
\end{subfigure}
\begin{subfigure}[b]{0.5\linewidth}
	\centering
	\includegraphics[width=0.95\linewidth]{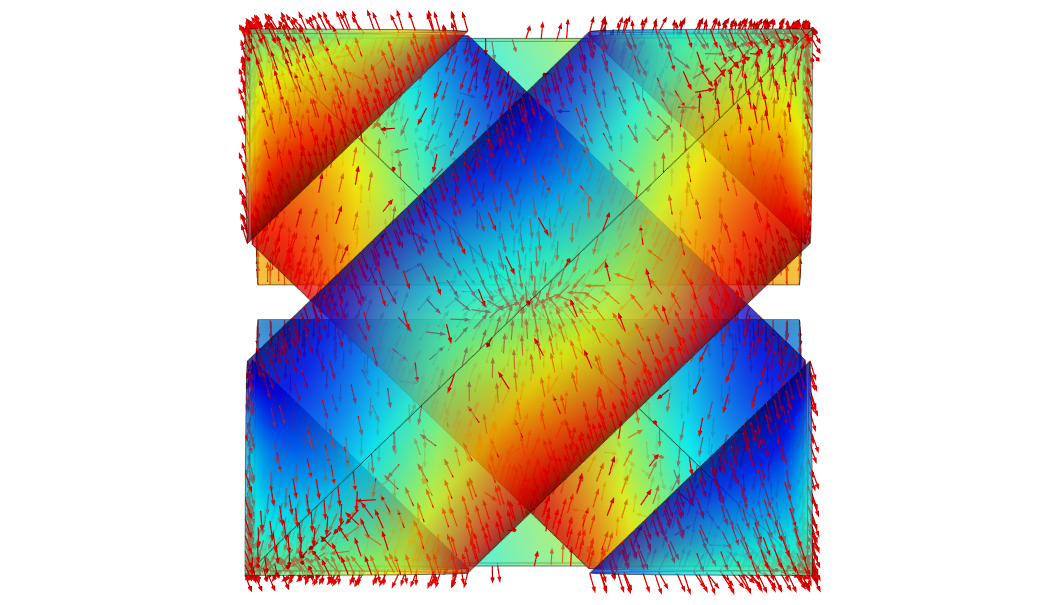}
	\caption{$\mathbf{\chi}^{22}$}
\end{subfigure}
\begin{subfigure}[b]{0.5\linewidth}
	\centering
	\includegraphics[width=0.95\linewidth]{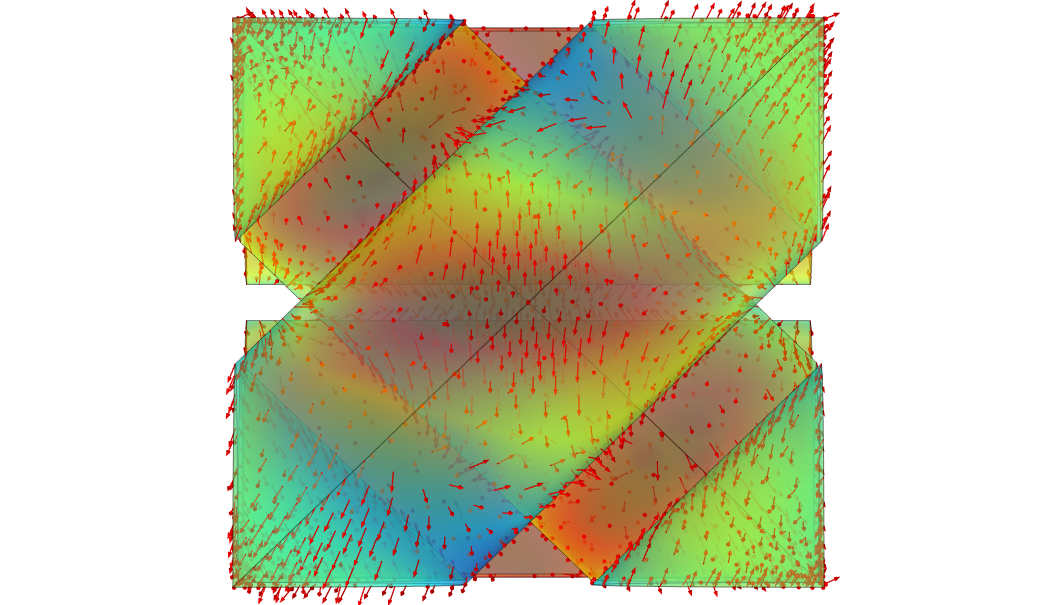}
	\caption{$\mathbf{\chi}^{33}$}
\end{subfigure}
\begin{subfigure}[b]{0.5\linewidth}
	\centering
	\includegraphics[width=0.95\linewidth]{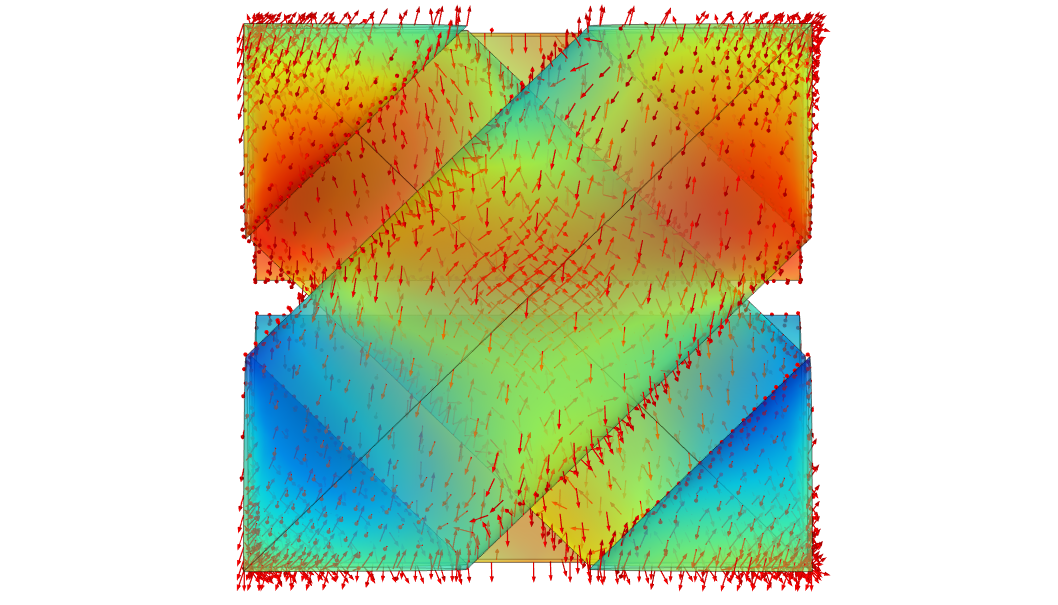}
	\caption{$\mathbf{\chi}^{23}$}
\end{subfigure}
\begin{subfigure}[b]{0.5\linewidth}
	\centering
	\includegraphics[width=0.95\linewidth]{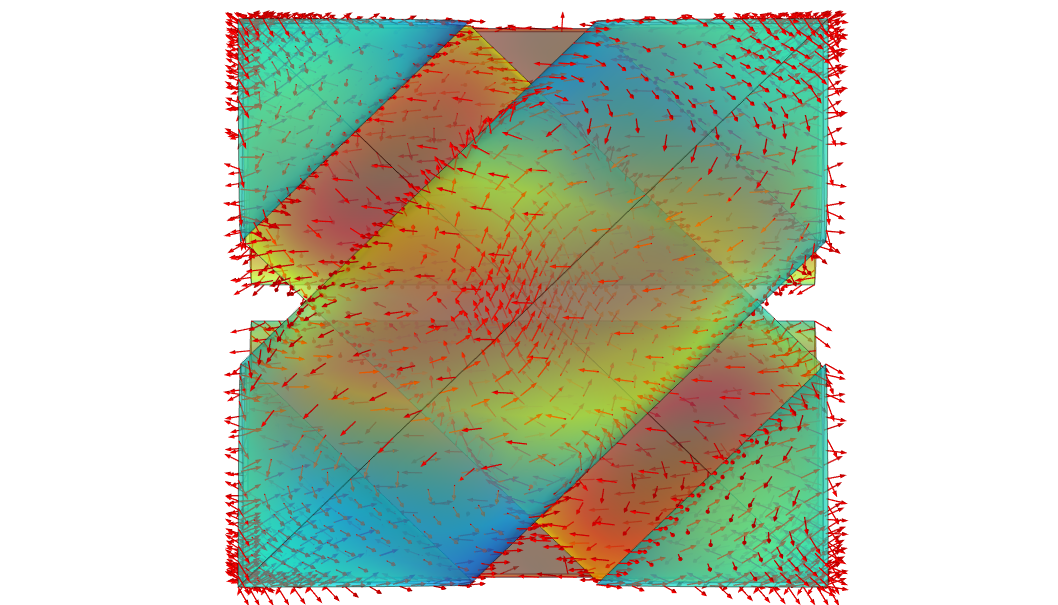}
	\caption{$\mathbf{\chi}^{13}$}
\end{subfigure}
\begin{subfigure}[b]{0.5\linewidth}
	\centering
	\includegraphics[width=0.95\linewidth]{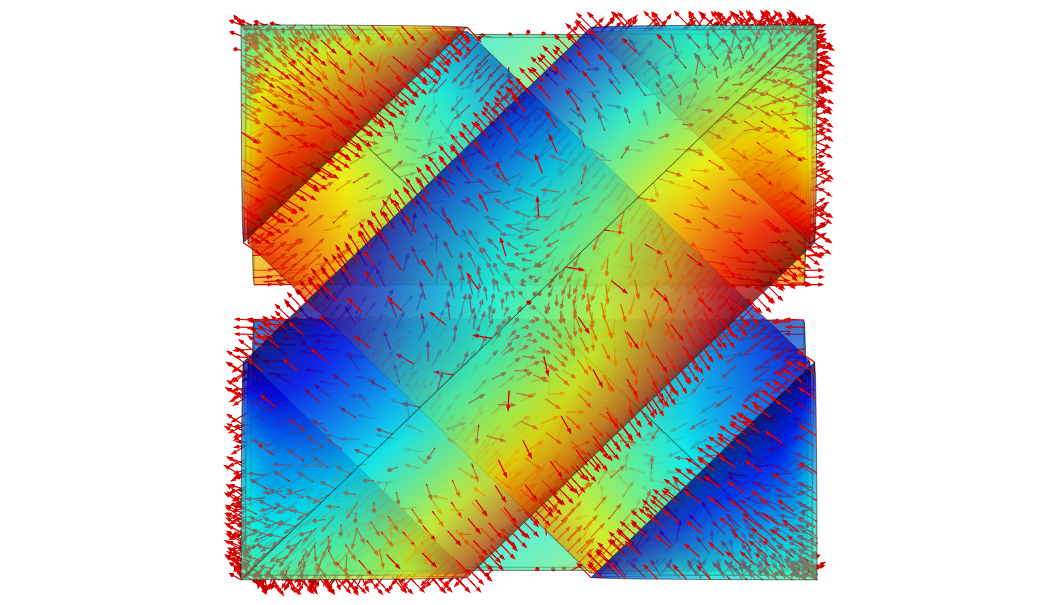}
	\caption{$\mathbf{\chi}^{12}$}
\end{subfigure}
	\caption{A visual account of the corrector function for a composite with $\phi=45^{\circ}$. The correctors are plotted only on the surface of the fibers and the top layer of $0^\circ$ fibers are hidden. The color corresponds to $(\chi^{mn}_1)^2 +  (\chi^{mn}_2)^2+(\chi^{mn}_3)^2$. Although not easily visible due to 3D plotting, the solution for the corrector is periodic.}
	\label{fig:correctors}
\end{figure}

$V_{f_1}$ corresponds to the volume fraction of the $0^\circ$ fibers and $V_{f_2}$ the volume fraction of each of the layers of the $\pm 45^{\circ}$ fibers.
Correspondingly, $a_1,a_2$ are the major radii of the elliptical cross section of the fibers and $b_1, b_2$ the minor radii (\autoref{fig:parametrization}).
A uniform distribution is considered for the aforementioned parameters, in the ranges presented in \autoref{tab:geomprops}.
In the current study, 200 model runs were used, with input vectors randomly sampled with Latin Hypercube Sampling (LHS) in order to explore the parameter space as 
well as possible with the limited budget of model runs. A visual account of the solution for the corrector function for a particular set of parameters is given in \autoref{fig:correctors}.

\subsection{Dimensionality reduction with PCA for the homogenized stiffness tensor}
It is natural to expect that the components of the homogenized stiffness tensor co-vary. 
In general, for arbitrary micro-structure geometries it is not trivial to assess intuitively the effect of geometric variation 
on the stiffness tensor directly. In the present study, Principal Component Analysis (PCA) is implemented
for the reduction of the $6\times6$ homogenized random stiffness tensor. In order to apply PCA on the homogenized tensors, 
the components of every random tensor are first flattened to a row vector as indicated in \autoref{eq:flattening}.

A set of  $N^{PCA}$ principal components $D^{(m)}_{PCA}$ (corresponding to tensor components) is sought, that satisfy \autoref{eq:pca},
where $\mu_{D^h} $ is the empirical mean of the homogenized stiffness, $\mathbf{X}_n$ is the
$n^{th}$ realization of the random input vector and $\lambda^{(m)}(\mathbf{X}_n)$ denotes a random coefficient that 
depends on the $n^{th}$ realization of the random input data.
\begin{equation}
	\label{eq:flattening}
	D^h_{ijkl}(\mathbf{X}_i) = {
	\begin{bmatrix}
		D_{1111} & D_{1122} & D_{1133} & 0        & 0        & 0 \\
		         & D_{2222} & D_{2233} & 0        & 0        & 0 \\
		         &          & D_{3333} & 0        & 0        & 0 \\
		         &          &          & D_{2323} & 0        & 0 \\
		         & Sym         &          &          & D_{1313} & 0 \\
		         &          &          &          &          & D_{1212} \\
	\end{bmatrix}
		} \rightarrow {\begin{Bmatrix}   D_{1111}\\ D_{2222} \\ D_{3333} \\ D_{2323} \\ D_{1313} \\ D_{1212} \\ D_{2233} \\ D_{1133} \\ D_{1122} \end{Bmatrix}^T  }
\end{equation}
\begin{equation}
	\label{eq:pca}
	D^h_{ijkl}(\mathbf{X}_n) = \sum_{m=1}^{N^{PCA}} \lambda^{(m)}(\mathbf{X}_n) \cdot D^{(m)}_{PCA}+\mu_{D^h}
\end{equation}

For the present study, the tensor is symmetric and it is expected to correspond to an ortho-tropic elastic material.
This results in 9 non-zero components.
For a general anisotropic elastic material, up to 21 components would be expected. Polynomial surrogates and sensitivity 
analysis for generally anisotropic materials described by probabilistically 
modelled random materials and random geometry may be treated via the same framework 
in a straightforward manner without placing any assumptions on the form of the stiffness tensor.
The $4$ first principal components were employed herein. Their contributions to the variance of the data are summarized in table \autoref{tab:varexplained}.
Considering the variance explained, it is concluded that 4 components are sufficient to capture the main 
variations on the homogenization data. 
The variance due to the remaining 5 principal components is considered insignificant, and attributed to the slight inaccuracies of the FE solution.
For illustrative purposes, the two first principal components are presented in \autoref{tab:components}.
\begin{table}[!h]
\centering
\begin{tabular}{l | c | c | c| c  }
	Component & $D^{(1)}_{PCA}$ & $D^{(2)}_{PCA}$ &  $D^{(3)}_{PCA}$ & $D^{(4)}_{PCA}$ \\
	\hline
	Explained Variance   & 50.01\% & 35.45\% & 14.10\% & 0.39 \% \\
\end{tabular}
\caption{Variance explained by the first 4 principal components.}
\label{tab:varexplained}
\end{table}
\begin{table}

\centering

\begin{align*}
	D^{(1)}_{PCA} & = 
	{\begin{bmatrix}
0.60&0.11&0.05&0 & 0 &0 \\
0.11&0.46&0.05&0&0&0\\
0.05&0.05&0.16&0&0&0\\
0&0&0&0.25&0&0\\
0&0&0&0&0.27&0\\
0&0&0&0&0&0.50\\
	\end{bmatrix}} \\
	D^{(2)}_{PCA} & = 
	{\begin{bmatrix}
0.74&-0.12&-0.01&0&0&0\\
-0.12&-0.14&-0.02&0&0&0\\
-0.01&-0.02&-0.04&0&0&0\\
0&0&0&-0.10&0&0\\
0&0&0&0&-0.04&0\\
0&0&0&0&0&-0.64\\
\end{bmatrix}} 
\end{align*}
\caption{First two principal components of the random homogenized stiffness tensor.}
\label{tab:components}
\end{table}

In what follows, polynomial chaos expansions and Sobol' sensitivity 
analysis are implemented on the coefficients of the 4 principal components.
Polynomial chaos expansion is constructed from a tensor product basis of polynomials orthogonal with respect to the probability distribution of the random 
inputs of our problem. In the present problem, since the distribution of all input random variables is uniform, 
a basis composed of multivariate tensor products of \textit{Legendre} polynomials in each of the input variables of \autoref{tab:geomprops} si employed.
Namely, polynomial chaos expansion is sought in the form
\begin{equation}
	\hat{{D}}^h_{PCE}(\mathbf{X}) = \sum_{m=1}^{N^{PCA}} \sum_{\mathbf{a} \in \mathcal{A}} c^{(m)}_{\mathbf{a}} \mathbf{\Psi}_{\mathbf{a}}(\mathbf{X}) D^{(m)}_{PCA}
\end{equation}
with $\mathbf{X} = \{ x_1,\cdots,x_6 \}$ denoting the random parameters of the micro-structure,  ${D}^{(m)}_{PCA} $ and $m=\{1,2,\cdots,N^{PCA}\} $ denoting the 
principal components. In our case $N^{PCA} = 4 $.

The linear PCA approach adopted herein straightforwardly allows for the approximate reconstruction of the 
original stiffness tensors. The same approach was assessed in the context of health 
monitoring in \cite{yunus}.

\begin{figure}[!h]
	\centering
	\includegraphics[width=0.5\textwidth]{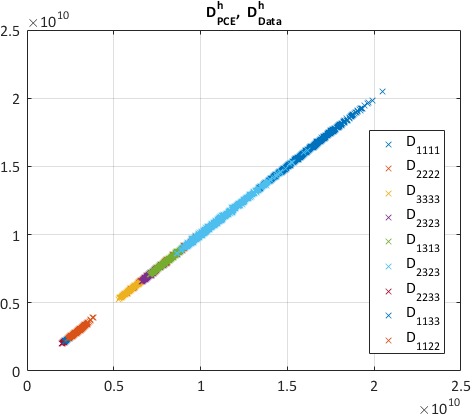}
	\caption{Quality of fit for the stiffness tensor. The tensor components are retrieved by using the PCE approximated PCA component coefficients with the principal component vectors.}
	\label{fig:PCEfit}
\end{figure}

The polynomial chaos expansion is computed by means of Least Angle Regression (LAR) \cite{blatmanSudret}.
The quality of the PCE least angle regression fit is measured with the generalized LOO 
error \cite{friedmanBook}. In \autoref{tab:fitqual} various 
parameters indicative of the quality of the PCE regression fit are summarized, 
separately for different principal components of the homogenized tensor.
A visual account of the quality of the fit for all reconstructed stiffness tensor components, is
demonstrated by plotting the reconstructed components against the original 
simulation data in \autoref{fig:PCEfit}.

\begin{table}
	\centering
\begin{tabular}{c | c | c |c}
	Component & {PCE-LOO Error} & Normalized MSE & PCE Maximum  Degree  \\
	\hline
$\lambda^{(1)}$ & $1.90e-3 $  &  $1.6e-3$           &   $ 6 $            \\
$\lambda^{(2)}$  & $2.05e-3 $  &  $1.7e-3$           &   $ 5 $            \\
$\lambda^{(3)}$  & $1.81e-3 $  &  $1.3e-3$           &   $ 5 $            \\
$\lambda^{(4)}$  & $7.17e-2 $  &  $5.0e-2$             &   $ 5 $            \\
\end{tabular}
	\caption{PCE least-squares regression fit quality measures and maximum degree of expansion for the principal components with LAR.}
	\label{tab:fitqual}
\end{table}

\begin{figure}
	\centering
	\includegraphics[width=0.95\textwidth]{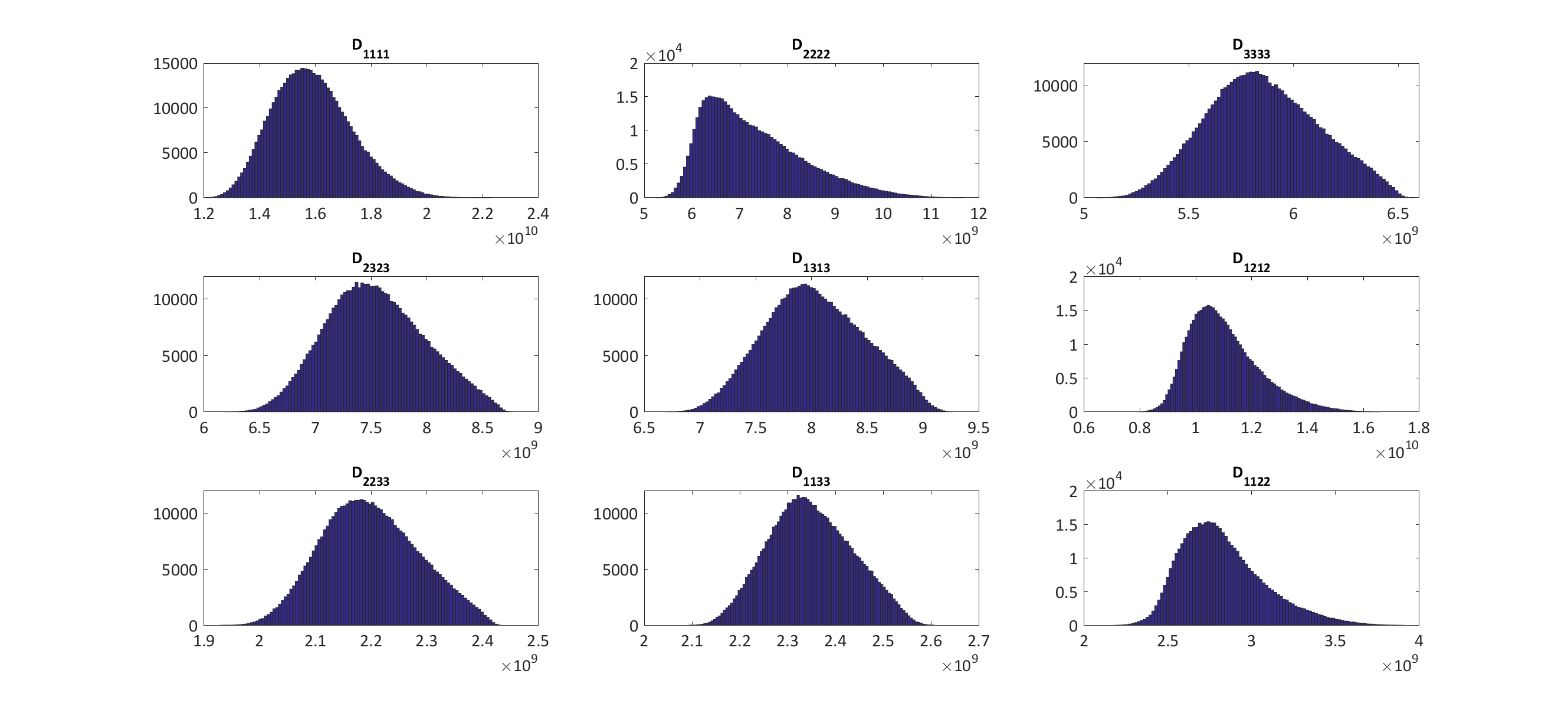}
	\caption{Histogram of coefficients of the homogenized stiffness matrix for the selected variation of parameters.}
	\label{fig:histograms}
\end{figure}

The performance of the fit is considered as satisfactory. In the next section the effect of the variability of the input variables to the homogenized stiffness is to be quantified by means of Sobol' sensitivity indices.
A set of histograms for the stiffness matrix component coefficients is given in \autoref{fig:histograms}. These histograms were computed by sampling from the polynomial chaos surrogate with $10^4$ samples.

\subsection{Sobol' Sensitivity Analysis}
\begin{figure}[!h]
\begin{subfigure}[b]{0.5\linewidth}
	\centering
	\includegraphics[width=0.95\linewidth]{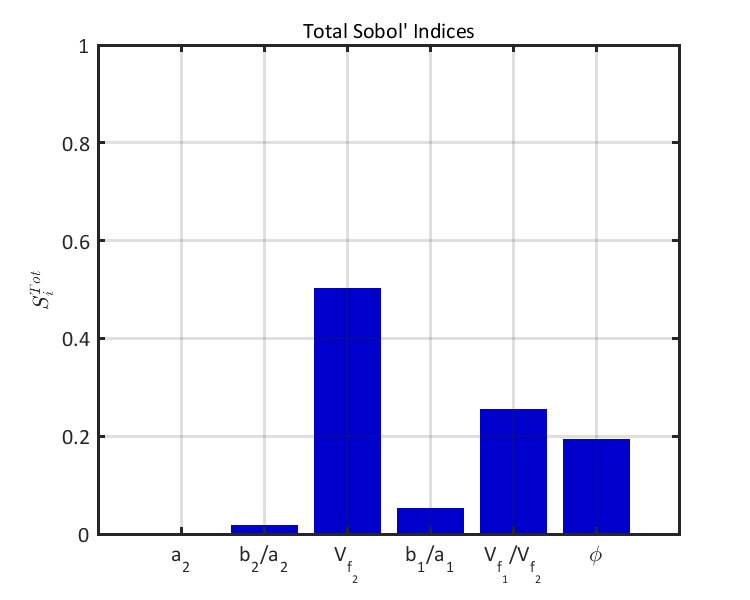}
	\caption{$\lambda^{(1)}$}
\end{subfigure}
\begin{subfigure}[b]{0.5\linewidth}
	\centering
	\includegraphics[width=0.95\linewidth]{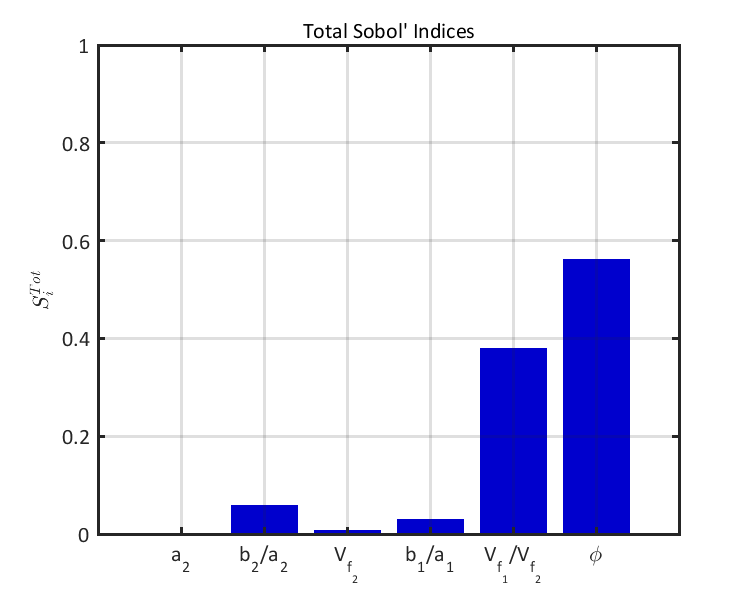}
	\caption{$\lambda^{(2)}$}
\end{subfigure}
\begin{subfigure}[b]{0.5\linewidth}
	\centering
	\includegraphics[width=0.95\linewidth]{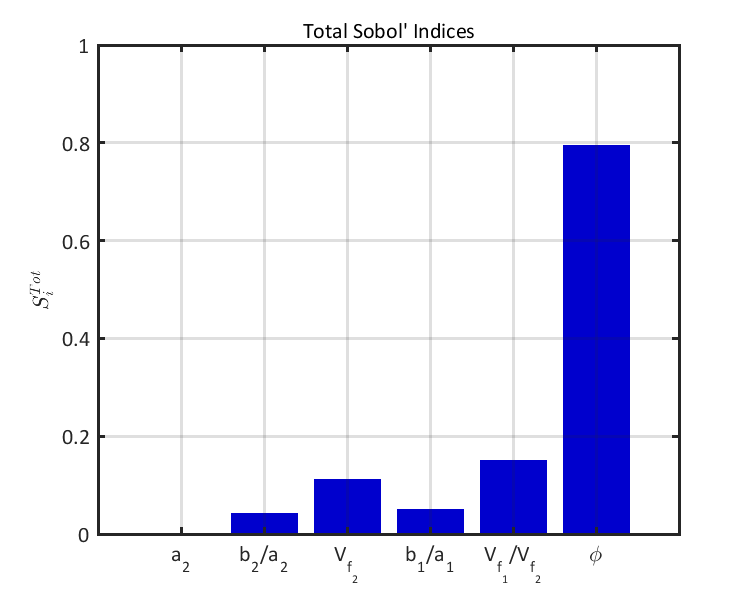}
	\caption{$\lambda^{(3)}$}
\end{subfigure}
\begin{subfigure}[b]{0.5\linewidth}
	\centering
	\includegraphics[width=0.95\linewidth]{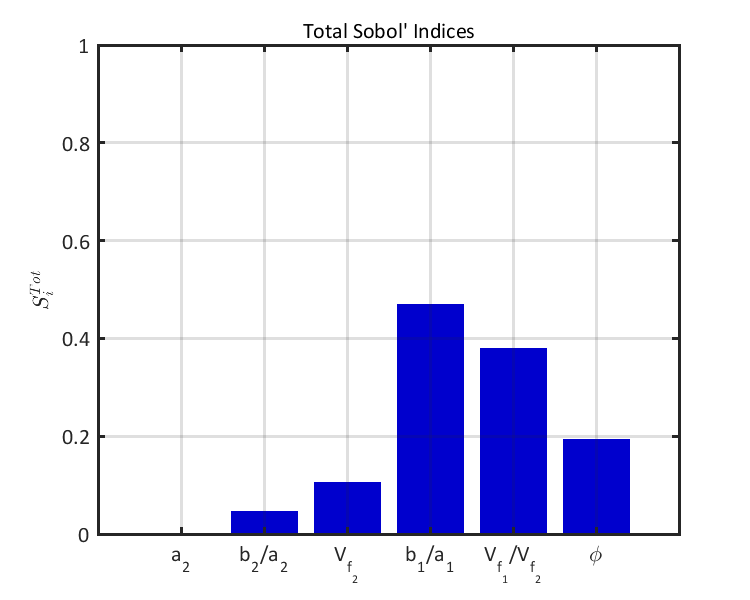}
	\caption{$\lambda^{(4)}$}
\end{subfigure}
	\caption{Sobol' sensitivity indices for the different components of the PCA of the homogenized tensor data.}
	\label{fig:totsobol}
\end{figure}

As demonstrated in \cite{sudret2008} it is possible to efficiently compute the Sobol' global sensitivity indices through the coefficients of a polynomial chaos surrogate model.
Sensitivity analysis is performed separately for each on of the 4 principal components of the PCA.
The results are presented in \autoref{fig:totsobol}.
It should be noted that the results of the sensitivity analysis
on the $\lambda^{(m)} $ have a meaning that is not decoupled from 
the values of the principal components $ D^{(m)}_{PCA} $ themselves (\autoref{tab:components}).
In a setting where the principal components had an interpretable meaning such an analysis 
would have been more beneficial.

Nevertheless, in our setting, it is clear that the angle of the $\pm \phi^\circ$ fibers is a significant factor, along with $V_{f_2}$ and the ratio of the volume fractions of  $\pm \phi^\circ$ and $0^\circ$ fibers.
It is interesting to observe, that the shape of the fibers, represented by  $a_1,a_2,b_1,b_2$ has an almost negligible effect on the homogenization problem, at least for the range of variations considered in the 
present study. The high sensitivity index in the $4^{th}$ principal component is considered negligible, in light of the small contribution to the variance in the context of PCA of $\lambda^{(4)}$.

\subsection{Conclusion}

A framework for the construction of 
efficient and accurate polynomial surrogate models 
is presented for the problem of homogenization of parametrized, probabilistically 
modelled random microstructures. A limited budget of 
random Monte-Carlo runs is employed together with 
a non-intrusive surrogate modelling approach.
Linear Principal Component Analysis was found sufficient 
for the data-driven dimensionality reduction of the 
random realizations of the stiffness tensor.
Efficient PCE-based global sensitivity analysis was performed, 
yielding quantitative results on the effect of different 
random input parameters on the composite macro-scale response.

The utility of PCE models for 
the homogenization and localization
problems is not limited to the gaining of a deeper insight 
on the effect of uncertainty of input parameters on homogenization through sensitivity analysis,
as demonstrated in the present study. 
Although in the present work homogenization 
surrogates are exclusively presented, a rather simple extension in the same 
framework would pertain to the construction of surrogate 
models for the problem of micro-strain computation 
under uncertainty. This will form part of future investigations.
The efficient solution of the stress localization problem efficiently is an important stepping 
stone towards the goal of highly efficient multi-scale damage 
prediction for composites of an arbitrary micro-structure.

\paragraph{Acknowledgement:}
The authors would like to gratefully acknowledge the support of the European Research Council via the ERC Starting Grant WINDMIL 
(ERC-2015-StG \#679843) on the topic of  Smart Monitoring, Inspection and Life-Cycle Assessment of Wind Turbines.

\clearpage

\bibliography{biblio}

\bibliographystyle{unsrt}



%
\end{document}